\documentclass[12pt]{article}

\usepackage{latexsym}
\usepackage{graphicx}\usepackage{epsfig}

\usepackage{epsf}
\usepackage{longtable}
\usepackage{amssymb,amsmath,amsfonts}

\newcommand{\beq}{\begin{eqnarray}}
\newcommand{\eeq}{\end{eqnarray}}

\newcommand{\ie}{{\it i.e.,}\ }

\begin {document}

\thispagestyle{empty}

\begin{titlepage}
\begin{flushright}
{\small
}
\end{flushright}
\vskip 1cm

\centerline{\Large \bf Exact Microscopic Entropy of Non-Supersymmetric}
\vskip 0.3cm
\centerline{\Large \bf  Extremal Black Rings}

\vspace{2.5cm}

\centerline{
Roberto Emparan}

\vskip .5cm
\centerline{\em Instituci\'o Catalana de Recerca i Estudis
Avan\c cats (ICREA),} 
\centerline{\em and}
\centerline{\em Departament de F{\'\i}sica Fonamental, Universitat de
Barcelona,}
\centerline{\em Marti i Franqu{\`e}s 1,
E-08028 Barcelona, Spain}

\vskip .7cm

\centerline{\tt emparan@ub.edu}

\vskip 1cm

\begin{abstract}
 
\noindent In this brief note we show that the horizon entropy of the
largest known class of non-supersymmetric extremal black rings, with up
to six parameters, is exactly reproduced for all values of the ring
radius using the same conformal field theory of the four-charge
four-dimensional black hole. A particularly simple case is a dipole
black ring without any conserved charges. The mass gets renormalized,
but the first corrections it receives can be easily understood as an
interaction potential energy. Finally, we stress that even if the
entropy is correctly reproduced, this only implies that one sector of
chiral excitations has been identified, but an understanding of excitations in the other
sector is still required in order to capture the black ring dynamics.

\end{abstract}
 
\end{titlepage}

\setcounter{footnote}{0}

\paragraph{Motivation.} 

Non-supersymmetric extremal rotating black holes provide a convenient
setting for furthering our microscopic understanding of quantum
processes related to Hawking radiation. One the one hand, they exhibit
the attractor behavior that allows precise control over the statistical
counting of the entropy. On the other hand, they typically possess
ergoregions and thus emit quantum radiation in a process closely related
to Hawking radiation, even if they are at zero temperature
\cite{Page:1976ki}: this is the spontaneous superradiant emission
recently studied in a microscopic model in \cite{Dias:2007nj}. Thus,
these black holes allow rather detailed control over their microscopic
description, while exhibiting richer and more realistic dynamics than
supersymmetric black holes.

The non-supersymmetric extremal black {\it rings}
\cite{Emparan:2004wy,Elvang:2004xi,Emparan:2006mm} that we discuss here
present further advantages in these respects over other non-BPS extremal
black holes. The ratio $R_2/R_1$ between the $S^2$ and $S^1$ radii of
the ring controls the departure from the BPS limit, which is recovered
as $R_1\to \infty$. Even if the mass is renormalized as a function of
this parameter, we shall argue that in fact such a renormalization can
be understood in simple terms in an expansion in $R_2/R_1$. In contrast,
the entropy is exactly reproduced for {\it all} values of $R_2/R_1$.
This result matches nicely with the attractor behavior discussed in
\cite{Astefanesei:2006dd,Goldstein:2007km} and provides a non-trivial
test of it. Very recently, the exact entropy of a vacuum black ring,
with horizon topology $S^1\times S^2$ and maximal rotation on the $S^2$
\cite{Reall:2007jv}, has been reproduced following on ideas in
\cite{Emparan:2006it,Emparan:2007en,Horowitz:2007xq}. However, in this
case the connection between the dual CFT and the black hole degrees of
freedom is obscured by the series of transformations involved in the
mapping. In our examples, the connection to the CFT is much more direct.
Still, they present a number of novel features, yet to be fully
understood, largely due to the fact that, in contrast to supersymmetric
black holes and black rings, these black rings can emit quantum
radiation.

Thus we believe that understanding in detail the microscopic dual of
these black rings can significantly enlarge the range of black hole
phenomena we can address using string theory. As a first step in the
correct identification of this dual description, in this note we show
how their entropy is exactly reproduced by a microstate counting.

\paragraph{Extremal dipole black rings.} Our black rings appear at the
common intersection of three stacks of M5 branes over a circle. We begin
with the simplest example of these, constructed in
\cite{Emparan:2004wy}, which does not possess any conserved charges, only
three dipoles $q_i$ associated to each kind of M5 brane. Thus the
solutions can never be supersymmetric, but they admit an extremal
(zero-temperature) limit with a regular horizon of finite area.
Ref.~\cite{Emparan:2004wy} proposed to account for their entropy using
the MSW conformal field theory \cite{Maldacena:1997de}, and indeed found
perfect agreement for the entropy up to the next-to-leading order in an
expansion in $R_2/R_1$. Although this is already a non-trivial matching,
ref.~\cite{Emparan:2004wy} failed to realize that the matching is exact
for arbitrary values of $R_2/R_1$, as we shall presently see. Later, the
idea to apply the MSW CFT to black rings was shown to work for
supersymmetric black rings \cite{Elvang:2004rt,Bena:2004tk,Cyrier:2004hj}. However,
these calculations left a number of puzzling issues; we will discuss
some of them after we deal with the more general solutions of
\cite{Elvang:2004xi}, which have M2 charges.

The extremal dipole black rings with three M5 dipoles are recovered from
the solutions in \cite{Emparan:2004wy} by setting the parameter $\nu$ to zero. The
solutions are expressed in terms of four dimensionless parameters
$\lambda$, $\mu_{1,2,3}$, with $0<\lambda,\mu_i<1$. Mechanical
equilibrium of the ring then determines that
\beq\label{equil}
\lambda=\frac{\mu_1+\mu_2+\mu_3+\mu_1\mu_2\mu_3}{1+\mu_1\mu_2+\mu_1\mu_3+\mu_2\mu_3}
\,.
\eeq
The solutions also contain a scale $R$. Roughly, $R$ measures
the size of the ring's $S^1$ radius, $R\sim R_1$; $\lambda$ is roughly the ratio
between the $S^2$ and $S^1$ radii, \ie $R_2\sim \lambda R$. Finally, the
$\mu_i$ are ratios between
`dipole-charge radii' and $R$. A straight string is recovered as
$\lambda,\mu_i\to 0$ keeping $\lambda R$ and $\mu_i R$ finite. 

The Bekenstein-Hawking horizon
entropy is 
\beq\label{bhentropy}
S_{BH}=\left(\frac{2\pi R^3}{G}\right)
\pi\sqrt{\lambda(1-\lambda^2)}
\prod_{l=1}^3\mu_l^{1/2}(1+\mu_l)\,,
\eeq
and the integer-quantized M5 dipoles are 
\beq
n_i=\epsilon_i\left(\frac{2\pi R^3}{G}\right)^{1/3}
\mu_i\frac{\sqrt{1-\lambda}}{\sqrt{1-\mu_i^2}}
\prod_{l=1}^3\sqrt{1+\mu_l}\,.
\eeq
The rings also possess an angular momentum along the $S^1$ direction
\beq
J_1=\epsilon_\lambda\frac{1}{4}\left(\frac{2\pi R^3}{G}\right)
\lambda\sqrt{1+\lambda}\prod_{l=1}^3(1+\mu_l)^{3/2}\,.
\eeq
We have included a choice of sign $\epsilon_\lambda,\epsilon_i=\pm
1$ for later generality. For our purposes at finite $R_2/R_1$ (\ie finite
$\lambda$, $\mu_i$) the orientations of the momentum and the M5 branes
may remain arbitrary. It is
only if we want to recover a supersymmetric solution in the
straight-string limit that we must correctly align the momentum and branes. 

A direct calculation shows that, for {\em all} allowed values of the
parameters, the entropy \eqref{bhentropy} can be written as
\beq\label{dipoleS}
S_{BH}&=&2\pi\sqrt{|J_1 n_1 n_2 n_3|}\,.
\eeq
One naturally expects that if we view the ring as a circular string,
then $J_1$ must be identified as the units $n_p$ of linear
momentum running along the string---detailed arguments to this effect were given in
\cite{Emparan:2004wy} and \cite{Emparan:2005jk}. Then \eqref{dipoleS}
is exactly the Cardy entropy 
\beq\label{cardyS}
S_{CFT}=2\pi\sqrt{\frac{c |\hat q_0|}{6}}
\eeq
that the MSW conformal field theory, with central charge
\beq\label{ccharge}
c=6 |n_1 n_2 n_3|\,,
\eeq
assigns to a state
where
one of the chiral sectors of the theory carries momentum
\beq
|\hat q_0|=|n_p|=|J_1|\,.
\eeq

The mass of the black ring is however different than the naive mass for
the MSW string: it is renormalized as one increases the ratio $R_2/R_1$.
Still, we can easily understand this effect in an expansion
for large $R_1$, which corresponds to small $\lambda,\mu_i$. If we
take into account \eqref{equil} we find
\beq
M=\frac{\pi R^2}{4G}\left( 2\lambda+2\sum_i\mu_i+
3\lambda^2-\sum_i\mu_i^2+O(\lambda^3,\mu_i^3)\right)\,.
\eeq
We can rewrite this in terms of the unit masses for M5 branes and momentum excitations,
\beq
m_5=R\left(\frac{\pi}{4G}\right)^{2/3}\,,\qquad m_p=\frac{1}{R} \,,
\eeq
so that
\beq\label{mass}
M=|n_p| m_p+\sum_i |n_i| m_5 -\frac{G\left((n_p m_p)^2+\sum_i (n_i
m_5)^2\right)}{\pi R^2}+O(1/R^3)\,.
\eeq
The leading $O(G^0)$ term is the BPS mass of the MSW string. The first
$O(G)$ correction,
which appears when the string is curved and supersymmetry is broken, is
the attractive potential energy due to the self-interaction among the
constituents. It shows the characteristic Newtonian fall-off $\propto 1/R^2$
of the five-dimensional potential energy.

\paragraph{Extremal non-BPS charged black rings}

Ref.~\cite{Elvang:2004xi} presented a larger class of extremal
non-supersymmetric black rings, which in addition to the three M5
dipoles possess conserved charges of three kinds of M2 branes, and
angular momenta $J_1$ and $J_2$ in the $S^1$ and $S^2$ of the ring.
Besides the parameters $R$, $\mu_i$, $\lambda$, which again must
satisfy \eqref{equil}, the solutions contain three boost parameters
$\alpha_i$ related to the M2 charges. They are constrained by
the condition 
\beq
  \frac{C_\lambda}{1+\lambda}
   {s_1}  {s_2}  {s_3}=
  \frac{C_1}{1-\mu_1}
   {s_1}  {c_2}  {c_3}
  +\frac{C_2}{1-\mu_2}
   {c_1}  {s_2}  {c_3}
  +\frac{C_3}{1-\mu_3}
   {c_1}  {c_2}  {s_3}
   \, 
  \label{DMcond}
\eeq
which derives from the absence of Dirac-Misner string
singularities, which give rise to global causal pathologies.
Here we have defined
\beq
C_\lambda=\epsilon_\lambda\lambda\sqrt{\frac{1+\lambda}{1-\lambda}}\,,\qquad 
C_i=\epsilon_i\mu_i\sqrt{\frac{1-\mu_i}{1+\mu_i}}\,,\qquad 
c_i= \cosh\alpha_i\,,\qquad s_i= \sinh\alpha_i\,,
\eeq
again with $\epsilon_\lambda,\epsilon_i=\pm
1$.
Thus we see that the solutions contain six independent continuous parameters. These
are {\it not} enough to allow independent variation of the eight
physical parameters of the extremal black rings, namely the M2 and M5
numbers $N_{1,2,3}$, $n_{1,2,3}$, and the two spins $J_{1,2}$ (the mass
is fixed at extremality, though not necessarily at the BPS bound). In
particular the BPS limit with three charges and three dipoles can never
be achieved within this family. The
more general solution that allows this freedom is still to be
constructed. 

The M2 and M5 brane numbers are
\beq
\label{Qs}
N_i=\left(\frac{2\pi R^3}{G}\right)^{2/3} \frac{s_i c_i}{2}
\Big[
  \lambda-\mu_i+\mu_j+\mu_k
  +2(\mu_j\mu_k+\lambda\mu_i)
  +\lambda(\mu_i\mu_j+\mu_i\mu_k-\mu_j\mu_k)+\mu_i\mu_j\mu_k
\Big]\;,
\eeq
\beq
 n_i&=&\left(\frac{2\pi R^3}{G}\right)^{1/3}\frac{\sqrt{1-\lambda}}{{s_i}}
    \Bigg[\prod_{l=1}^3 \sqrt{1+\mu_l}\Bigg]\left(
 \frac{C_j}{1-\mu_j} {s_j} {c_k}
 +\frac{C_k}{1-\mu_k} {c_j}  {s_k}\right)\;.\label{q3}
\eeq
Here $(ijk)$ are cyclic permutations of $(123)$. The angular momentum
along the $S^1$ of the ring is
\beq
J_{1}\!\!&=&\!\!\frac{1}{4}\left(\frac{2\pi R^3}{G}\right)
(1-\lambda)^{3/2}
\Bigg[
\prod_{l=1}^3(1+\mu_l)^{3/2}
\Bigg]\nonumber \\
  && \hspace{1cm}
  \times
\Bigg(\;
  \frac{C_{\lambda}}{1-\lambda}
   {c_1}  {c_2}  {c_3}
-\frac{C_1}{1+\mu_1} {c_1}  {s_2}  {s_3}
-\frac{C_2}{1+\mu_2} {s_1}  {c_2}  {s_3}
-\frac{C_3}{1+\mu_3} {s_1}  {s_2}  {c_3}
\;\Bigg) \,.
\eeq
The angular momentum $J_2$ on the $S^2$ is most simply given in terms of
the M2 and M5 charges
\beq
J_2=-\frac{1}{2}\left(n_1N_1+n_2N_2+n_3N_3\right)\,.
\label{JqQ}
\eeq
Finally, the horizon entropy is
\beq
  S_{BH} &=& \pi\left(\frac{2 \pi R^3}{G} \right) \,
  (1-\lambda)\lambda^{1/2}
  \Bigg[\prod_{i=1}^3 (1+\mu_i)\mu_i^{1/2}\Bigg]
   \nonumber \\
  && \hspace{1cm}
  \times
  \Bigg|
  \frac{C_\lambda}{\lambda} {c_1} {c_2} {c_3}
  +\frac{C_1}{\mu_1} {c_1} {s_2} {s_3}
  +\frac{C_2}{\mu_2} {s_1} {c_2} {s_3}
  +\frac{C_3}{\mu_3} {s_1} {s_2} {c_3}
  \Bigg| \, .
\eeq

After a very long and tedious calculation, we have found that this horizon
entropy admits a simple expression in terms of physical parameters,
\beq
S_{BH}=2\pi\sqrt{\left|-J_1 n_1 n_2 n_3 
-\frac{(N_1 n_1)^2 +(N_2 n_2)^2+(N_3 n_3)^2}{4}
+\frac{N_1 N_2 n_1 n_2 +N_1 N_3 n_1 n_3+N_2 N_3 n_2 n_3}{2}\right|} \,.
\eeq
To reproduce this expression using the Cardy entropy \eqref{cardyS} with
the central charge \eqref{ccharge}, now we take
level
\beq\label{hq0}
\hat q_0= -J_1
-\frac{(N_1 n_1)^2 +(N_2 n_2)^2+(N_3 n_3)^2}{4 n_1 n_2 n_3}
+\frac{1}{2}
\left(\frac{N_1 N_2}{n_3} +\frac{N_1 N_3}{n_2}+\frac{N_2 N_3}{n_1}\right)\,.
\eeq
Setting $n_p=-J_1$, the momentum in this level is shifted precisely by
the momentum excited by the worldvolume fluxes
that yield the M2 charges \cite{Maldacena:1997de}---this zero-mode momentum does not contribute
to the degeneracy of the state. A clear derivation of this result can be
found in \cite{Larsen:2005qr}\footnote{Ref.~\cite{Larsen:2005qr} also describes a class of
non-supersymmetric extremal black rings. In that case extremality is
achieved by spinning up the rotation of the $S^2$, so it appears to be a
different effect than in this paper, although a connection may still exist.}.

Thus the MSW theory naturally accounts for the entropy of these black
rings. The complexity of the equations, however, has prevented us from
obtaining an expression for the mass that extends \eqref{mass}.

However simple this application of the MSW theory looks, it deepens the
puzzles posed by previous application of essentially the same formulas
to BPS black rings. In \cite{Cyrier:2004hj} the entropy of the
supersymmetric ring was matched to an expression for $\hat q_0$
essentially like above, but with an additional zero-point contribution $+n_1n_2n_3/4$
to the momentum. In \cite{Bena:2004tk}, however, the entropy of {\it the
same} supersymmetric black ring was instead matched to an expression
without this zero-point term, but with $\mathcal{N}_i= N_i-n_jn_k$ instead of $N_i$.
Additionally, one should put $-(J_1+J_2)$ for the momentum units $n_p$,
instead of $-J_1$. None of these oddities appears in the expression we
have found for the extremal non-BPS ring, although it may be worth noting
that an alternative form for
the oscillator level, which also reproduces the correct entropy, is 
\beq
\hat q_0= -(J_1+J_2)
-\frac{(\mathcal{N}_1 n_1)^2 +(\mathcal{N}_2 n_2)^2+(\mathcal{N}_3 n_3)^2}{4 n_1 n_2 n_3}
+\frac{1}{2}
\left(\frac{\mathcal{N}_1 \mathcal{N}_2}{n_3} +\frac{\mathcal{N}_1 \mathcal{N}_3}{n_2}
+\frac{\mathcal{N}_2 \mathcal{N}_3}{n_1}\right)-\frac{3n_1n_2n_3}{4}\,.
\eeq

According to the proposal in \cite{Emparan:2005jk}, the correct
expression for $n_p$ should be the momentum associated to the vector
tangent to the $S^1$ at the horizon. This favors the choice
$n_p=-(J_1+J_2)$ in the BPS case (i.e., like in \cite{Bena:2004tk}), and it
also yields $n_p=-J_1$ for our extremal non-BPS rings\footnote{The claim
made in \cite{Emparan:2005jk} that this identification favors $n_p=-J_1$
for BPS rings is incorrect: the tangent to the $S^1$ of the
near-horizon-BTZ-black hole for these black rings is $\partial_\psi +
\partial_\phi$ since the coordinate $\phi$ on the $S^2$ is shifted by
$\psi$ at the horizon.\label{foot:btz}}. Still, if the
proposal of \cite{Bena:2004tk} is correct it remains to understand why
the M2 charges $N_i$ should be shifted in the BPS case but not in our
rings here. Presumably this has to do with the fact that in our extremal
non-BPS rings some modes are not excited, and in particular $J_2$
vanishes when the $N_i$ are zero. These puzzles would be greatly
clarified if we possessed the more general eight-parameter extremal
solution where $N_{1,2,3}$, $n_{1,2,3}$, $J_{1,2}$ can be varied
independently.

\paragraph{Final remarks.}

To finish, we comment on the fact that both the left and right
moving sectors of the CFT are crucial for understanding the state that
corresponds to a black ring, even if only one sector is needed to
account for the entropy.

Recall that the puzzles discussed above pertain only to black rings with
finite radius. As $R\to\infty$ the formulas of \cite{Bena:2004tk},
\cite{Cyrier:2004hj}, and of the present paper reduce to the same
expression, which accounts for the entropy of a supersymmetric black
string with M2 and M5 charges and momentum. In this limit, only the
non-supersymmetric sector of the CFT is excited.
This
identification of the degrees of freedom responsible for the entropy
captures only a part of the state, but not the fact that the ring wraps
a contractible circle.

Already for a supersymmetric state a difference exists between the cases
where the MSW string wraps a Kaluza-Klein circle and where it wraps a
(contractible) ring: the right-moving fermions in the former case are in
the R sector, while in the latter case they are in the NS sector. While it
would be interesting to study the consequences of this in more detail, a
more significant difference must exist between the BPS and extremal
non-BPS systems: the latter possess ergoregions and hence are unstable
to spontaneous superradiance. In \cite{BlancoPillado:2007iz} (see also
\cite{Emparan:2007en,Dias:2007nj}) this effect was explained as follows
(in the context of fundamental string states): {\it both} chiral sectors
of the theory are excited. One sector has a large degeneracy of states
and gives rise to the statistical entropy of the system. The other
sector is also excited (unlike in BPS systems), but in a coherent state
that gives the string a circular profile and does not contribute to the
degeneracy. Pairs of excitations from each of the two sectors can
interact and combine to form an outgoing quantum, which necessarily
carries angular momentum away. This accounts for spontaneous
superradiance. It seems likely that a similar picture should be present,
within the context of the MSW string, for the black rings we have
considered in this paper. So it should be interesting to understand the
excitations of the system that turn the string into a ring.

\paragraph*{Acknowledgements.} 
I am indebted to Harvey Reall for discussions that prompted me to
reconsider the microscopic calculation of ref.~\cite{Emparan:2004wy} and the
point discussed in footnote~\ref{foot:btz}. This
work was supported in part by DURSI 2005 SGR 00082, CICYT FPA
2004-04582-C02-02, and the European Community FP6 program
MRTN-CT-2004-005104.

\end{document}